\documentstyle[12pt,psfig]{article}

\begin{document}
\newcommand{\lash}[1]{\not\! #1 \,}
\newcommand{\bra}[1]{\big< #1 \big|}
\newcommand{\ket}[1]{\big| #1 \big>}
\def\be{\begin{eqnarray}}
\def\en{\end{eqnarray}}
\vskip 1.5 cm
\centerline{\large\bf Nuclear Shadowing Effects on Prompt Photons}
\centerline{\large\bf at RHIC and LHC}
\medskip
\bigskip
\medskip
\centerline{\bf N. Hammon, A. Dumitru, H. St\"ocker, W. Greiner}
\medskip
\bigskip
\centerline{Institut f\"ur  Theoretische Physik,}
\centerline{Robert-Mayer Str. 10,}
\centerline{Johann Wolfgang Goethe-Universit\"at,}
\centerline{60054 Frankfurt am Main, Germany}
\medskip
\bigskip
\medskip
\bigskip
\centerline{\bf Abstract}
\bigskip
{\small
The transverse momentum distribution of prompt photons coming from the
very early phase of ultrarelativistic heavy ion collisions for the
RHIC and LHC energies is calculated by means of perturbative QCD. 
We calculate the single photon cross section ($A+B\rightarrow \gamma +X$) by taking into 
account the partonic sub processes $q+ \overline{q}\rightarrow\gamma +g$ and 
$q+g\rightarrow\gamma +q$ as well as the Bremsstrahlung corrections to those processes.
We choose a lower momentum cut-off $k_0 =2$ GeV separating the soft physics from 
perturbative QCD. We compare the
results for those primary collisions with the photons 
produced in reactions of the thermalized secondary
particles, which are calculated within scaling hydrodynamics. The QCD processes are taken
in leading order. Nuclear shadowing corrections, which alter the involved 
nuclear structure functions are explicitly 
taken into account and compared to unshadowed results.
Employing the GRV parton distribution parametrizations we find that at RHIC prompt
QCD-photons dominate over the thermal radiation down to transverse momenta
$k_T \approx 2$ GeV. At LHC, however, thermal radiation from the QGP dominates
for photon transverse momenta $k_T \leq 5$ GeV, if nuclear shadowing effects on
prompt photon production are taken into account.\\
\\
\\
\\
Frankfurt, \today}

\newpage

\begin{figure}
\centerline{\psfig{figure=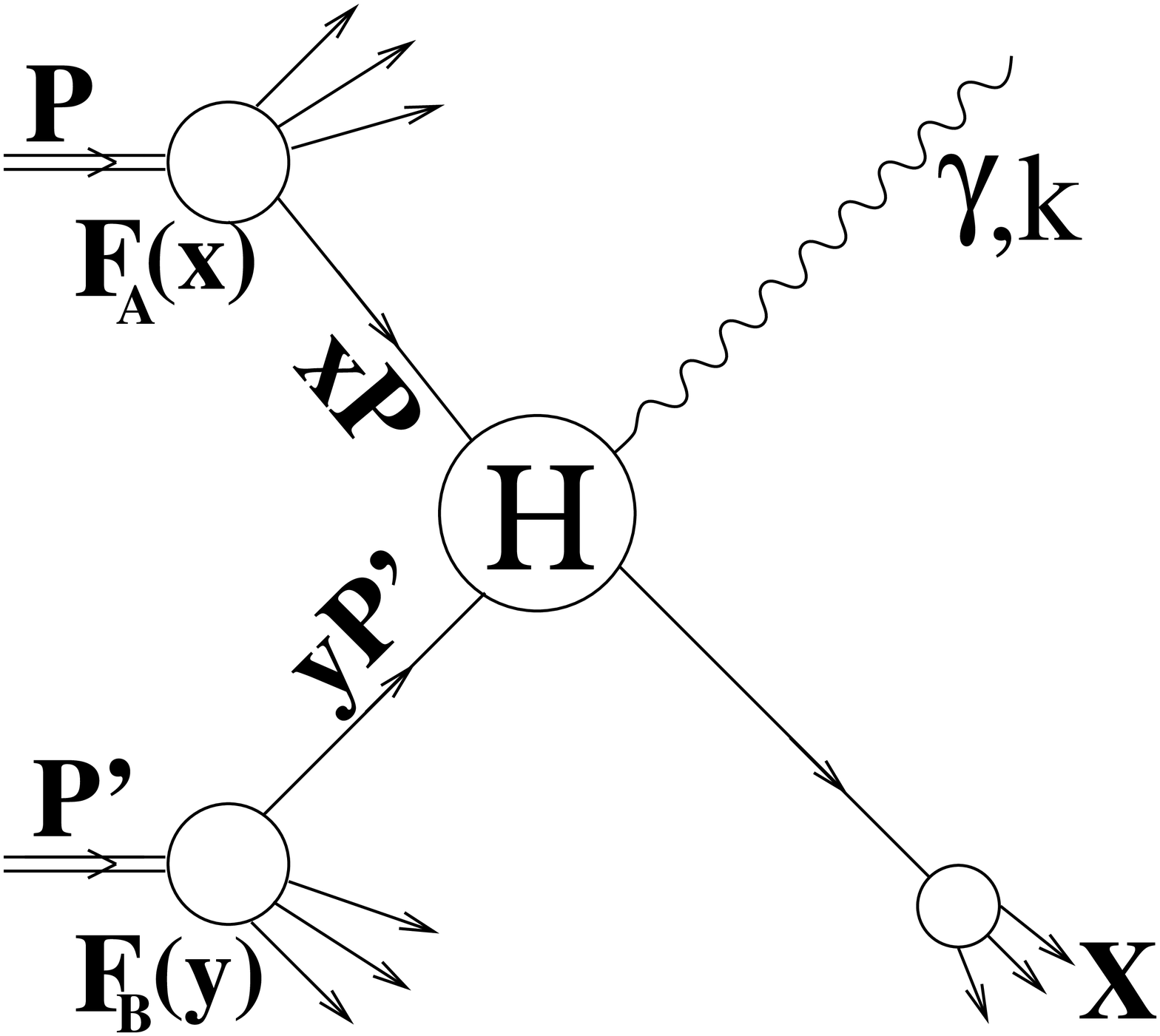,width=6cm}}
\caption[]{\sf General illustration of the direct single photon reaction. The
square of $F_A$ and $F_B$ generates the nucleon matrix element and $H$ is
the hard part of the reaction involving the different partonic subprocesses.}
\label{fig-general}
\end{figure}

\begin{figure}
\centerline{\psfig{figure=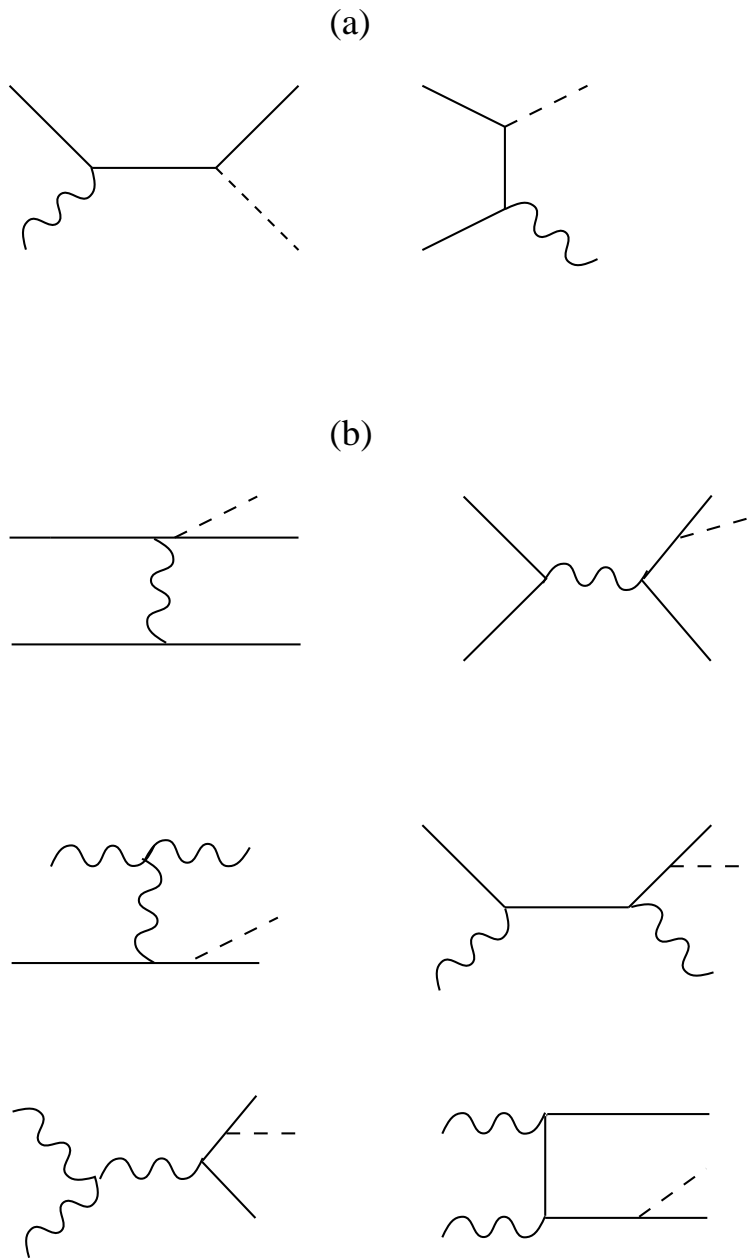,width=10cm}}
\caption[]{\sf (a) QCD Compton, annihilation and (b) Bremsstrahlung graphs taken into account in
our calculations. The wavy lines denote gluons and the dashed ones stand for the photons.}
\label{graphs}
\end{figure}

\begin{figure}
\centerline{\psfig{figure=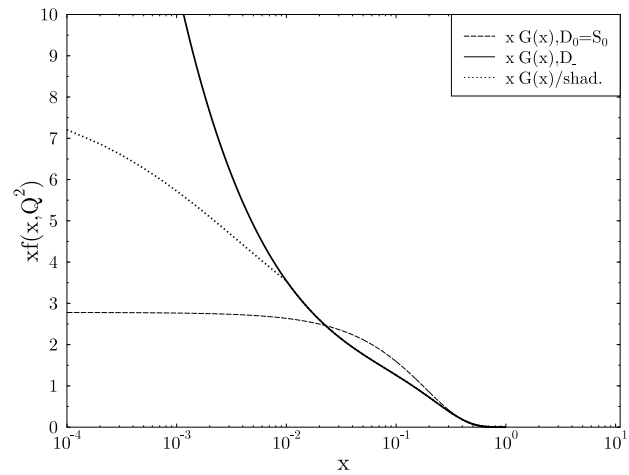,width=12cm}}
\caption[]{\sf Depletion of xG(x) in the proton as parametrized by MRS in \cite{MRS}.
The $D_{-}$ curve corrsponds to a singular behavior ($\sim x^{-0.5}$) as postulated by 
BFKL analysis whereas the small x behaviors of $S_{0}$ is taken from Regge anyalysis 
($\sim x^{0}$).}
\label{MRS-shad}
\end{figure}

\begin{figure}
\centerline{\psfig{figure=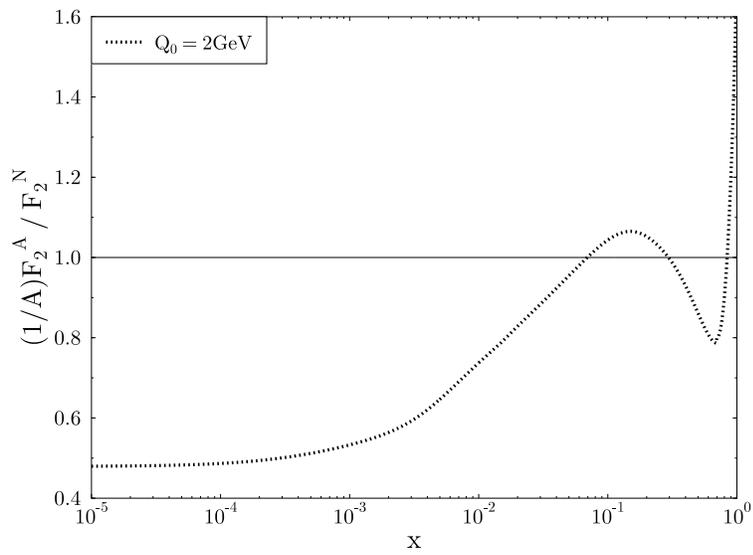,width=14cm}}
\caption[]{\sf Ratio $(1/A)F_2^A(x,Q^2)/F_2^P(x,Q^2)$ at the input scale $Q_0=2$ GeV
as parametrized in \cite{eskola}.}
\label{fig-kari}
\end{figure}

\centerline{\bf 1. Introduction}
In recent years a lot of effort has been made, on the experimental as well 
as on the theoretical side, to investigate the physics of the quark-gluon
plasma (QGP) \cite{qm96}. By doing so, one hopes to gain insight into the state of hot 
and dense matter created 
in a heavy ion collision and, in a more general manner, to learn about the 
evolution of the early universe which is believed to having passed this state 
shortly after the big bang.
During the time when the matter, produced 
in the collision of two
heavy nuclei at very high energies, is in the quark-gluon phase, particles
stemming from the interaction between the plasma constituents will arise.
By detecting the produced particles one hopes to gain knowledge of the physics
of the QGP. Those signatures for the plasma can be dilepton production,
$J/\psi$ suppression and photon production (for a review, see \cite{harris}).
It is important to know the contribution 
from collisions between the initial nucleons, i.e.~from a
non-plasma source, giving a background contribution to the thermal yield.\\
In this publication we consider prompt photon
production via the QCD processes $q+ \overline{q}\rightarrow\gamma +g$ and 
$q+g\rightarrow\gamma +q$ and by their Bremsstrahlung contributions up to 
${\cal O}(\alpha _ s ^2)$ at
midrapidity to investigate by which amount nuclear effects as shadowing
contribute to the hard process.
We take the formulas in leading-order and simulate the higher order terms by 
a K-factor.\\
In order to get
information about the primary $\gamma$'s one needs reliable information about 
the infrared dominated quantities entering the cross sections, i.e.~the parton 
distribution functions. The uncertainties appearing in connection with 
the QCD cross sections in nuclei are twofold. Not only one has to deal 
with the low-$x$ behavior of the parton distribution functions but also it is 
unavoidable to account for modifications of those distributions in nuclei, such as 
shadowing corrections, 
especially when considering heavy nuclei such as $Pb$ and $Au$.\\
We consider photon transverse momenta in the range 
$k_{T} = 2...5$ GeV which is relevant for the QGP (it was suggested \cite{shuryak}
that in the region $k_{T} = 2...4$ GeV the thermal photons from the QGP may dominate
whereas the hard QCD $\gamma$'s should dominate the larger $k_T$ region).\\
The hard collisions take place at some time $\tau \approx
1/k_0 \approx 0.1$ fm/c. The typical time scale for the QGP formation requires knowledge
of the dynamics of the formation and equilibration of the plasma. 
We therefore employ two sets of initial conditions for the temperature and the proper time in the hydro 
calculations to obtain a lower and an upper bound for the thermal yield.\\
\\
\\
\centerline{\bf 2. Inclusive single photon production in QCD}
We first consider the cross section of single $\gamma$ production on the
nucleon nucleon level.
The general form of a factorized inclusive cross section with large
momentum transfer $k_T$ can be written as \cite{qiu} 
\begin{equation}
\sigma (k_T) = H_0 \otimes f_2 \otimes f_2 +
(1/k_T) H_1 \otimes f_2 \otimes f_3 +(1/k_T^2)H_2 \otimes f_2 \otimes f_4
+{\cal O} (1/k_T^3)
\label{otimes}
\end{equation}
where the process at the nucleon nucleon level is 
\begin{equation}
N(P,S)+N'(P',S) \rightarrow \gamma (k) + X
\end{equation}
as illustrated in figure \ref{fig-general}.
%
%
The convolution is formulated in terms of the momentum fractions $x$ and
$y$. The $f_n$ are infrared dominated nonperturbative matrix elements of
twist $n$ and the $H_i$ are the perturbatively calculable coefficient
functions. 
In our approach we will work in LO with a K-factor for the higher orders in $\alpha _s$
and at the twist-2 level. The next twist to contribute in an unpolarized process
would be twist-4 which is suppressed by ${\cal O} (1/k_T^2)$ compared to leading twist
and is therefore neglected.
Matrix elements of twist-3 contribute only to polarized processes, such as single-spin
asymmetries \cite{qiu,nils}.\\
On the twist-2 level the cross section is
\begin{eqnarray}
E_{c} \frac{d\sigma}{d^3 k_{c}}(AB\rightarrow CX) =
\frac{1}{\pi}\sum\limits_{ab\rightarrow cd}^{}  
\int dx_{a}~dx_{b}~f_{A}^{a}(x_{a})~f_{B}^{b}(x_{b})\\ \nonumber
\hat{s}~\frac{d\hat{\sigma}}{d\hat{t}}(ab \rightarrow cd)
~\delta(\hat{s}+\hat{t}+\hat{u})
\end {eqnarray}
where the sum runs over all partonic subprocess including quark-antiquark annihilation into 
two photons and into photon and gluon and the QCD Compton process.
As stated above, we are here only interested in the {\it single} photon processes.
Reactions involving two final state photons, such as $q+\overline{q} \rightarrow \gamma 
+ \gamma$, are neglected since they are
suppressed due to the additional electromagnetic coupling constant.
Some comments on Bremsstrahlung will appear later because one gets a large contribution from those processes.\\
Since the above formula applys for the twist-2 level, the $f_{A,B}$ are
the well-known parton distribution functions of definite twist. For our calculations we 
used the newest 
version of the Gl\"uck, Reya and Vogt parametrisations \cite{GRV}.
The main reason for chosing the newest version is the modifications at small
$x$ measured at HERA \cite{HERA} which show a steep rise in the proton
structure function $F_2(x,Q^2)$. The photons at
midrapidity typically come from momentum fractions $x\approx 2k_0/\sqrt{s}$.
Therefore, at the high energy colliders treated here (RHIC and LHC) one clearly is in a region where
new data on parton distributions at low $x$ are important.
Another interesting property of the '95 GRV parametrisations
is the asymmetric sea taking into account the violation of the Gottfried sum
rule, which defines the difference between the proton and
neutron structure functions \cite{Gottfried}:
\begin{eqnarray}
I_{GSR}&=&\int_0^1 \frac{dx}{x}(F^p_2-F^n_2)\nonumber\\
&=&\frac{1}{3}\int_0^1 dx (u_v-d_v)+\frac{2}{3}\int_0^1 dx (\overline{u}
-\overline{d})=\frac{1}{3}~~{\rm if}~~\overline{u}=\overline{d}
\end{eqnarray}
The early analysis (prior to 1992) assumed a flavor
independent light-quark sea. The NMC collaboration, on the contrary, found \cite{NMC}
\begin{eqnarray}
&&\Sigma(0.004,0.8)=\nonumber\\
&&\int_{0.004}^{0.8}\frac{dx}{x}(F^p_2-F^n_2)=0.227\pm 0.007~
{\rm (stat.)}\pm 0.014 ~{\rm (syst.)}
\end{eqnarray}
implying $\overline{d} \ge \overline{u}$.\\
The (unpolarized) cross section for the prompt photon production 
in a nucleus nucleus reaction is given (without the exchange term) to 
lowest order by \cite{field}
\begin{eqnarray}
&&E_k\frac{d\sigma ^{AB}}{d^3 k}(A+B\rightarrow \gamma (k) +X)=\nonumber\\
&&\frac{1}{\pi}
\int_{x^a_{min}}^{1} dx_a f_i^A (x_a) f_j^B (x_b)
\frac{x_a x_b}{x_a -x_1} \frac{d\hat{\sigma}}{d\hat{t}}(a+b\rightarrow c+d)
\label{cross-section}
\end{eqnarray}
where $f_i^A (x_a)$ and $f_j^B (x_b)$ are the parton distributions {\it in the nucleus}.
The various hard scattering functions can e.g.~be found in \cite{alam} 
\begin{equation}
\frac{d\hat{\sigma}}{d\hat{t}}=\frac{\pi\alpha\alpha_s e_q^2}{\hat{s}^2} 
\frac{2}{3}\left[ \left(
\frac{\hat{s}}{-\hat{t}}\right)
+\left( \frac{-\hat{t}}{\hat{s}}\right)\right],
\end{equation}
\begin{equation}
\frac{d\hat{\sigma}}{d\hat{t}}=\frac{\pi\alpha\alpha_s e_q^2}{\hat{s}^2}
\frac{8}{9}\left[ \left(
\frac{\hat{u}}{\hat{t}}\right)
+\left( \frac{\hat{t}}{\hat{u}}\right)\right]
\end{equation}
for the processes $q+g\rightarrow\gamma +q$ and
$q+ \overline{q}\rightarrow\gamma +g$, respectively.\\
The partonic variables at midrapidity are given by
\begin{equation}
x_b = \frac{x_a x_T}{2x_a-x_t},x_T=2k_T/\sqrt{s},~x_a = x_b = k_T/\sqrt{s},~x_a^{min}
= \frac{x_T}{2-x_T}
\end{equation}
\begin{equation}
\hat{s}=x_a x_b s, \hat{u} = -x_b x_T s/2, \hat{t} = -x_a x_T s/2
\end{equation}
We also explicitely take the ${\cal O} (\alpha _s^2)$ corrections to the annihilation 
and Compton graphs, i.e.~Bremsstrahlung contributions, into account. 
Even though one has a ${\cal O} (\alpha _s^2)$ contribution, these processes already
contribute to ${\cal O} (\alpha _s)$
because the fragmentation function enters with a factor ln($Q^2/\Lambda ^2$).\\
In that process a photon is radiated off a final state quark, or in other words:
the quark fragments into a photon carrying a fraction $z$ of the initial momentum.
The fragmentation function is given in leading-log approximation by \cite{alam}
\begin{equation}
zD_{q\rightarrow \gamma}(z,Q^2)=e_q^2\frac{\alpha}{2\pi}[1+(1-z)^2] ~{\rm ln}(Q^2/\Lambda ^2)
\end{equation}
The result for the Bremsstrahlung contributions is given by \cite{alam}
\begin{eqnarray}
E_k \frac{d\sigma ^{AB}}{d^3 k} &=& 2\frac{\alpha \alpha _s^2}{2\pi s^2}{\rm ln}\frac{k_T^2}{\Lambda ^2}
\frac{1}{x_T}
\int_{x_T}^{1} dy_T \frac{4}{y_T^2}[1+(1-x_T/y_T)^2]\nonumber\\
& \times & \int_{y_T/(2-y_T)}^{1} \frac{dx_a}{x_a-y_T/2}
\Big [F_2 (x_a,A)\Big( xG(x_b,B)+\frac{4}{9}Q(x_b,B)\Big) \nonumber\\ 
& \times & \frac{x_a^2+(y_T/2)^2}{x_a^4}
+ (x_a\leftrightarrow x_b, A\leftrightarrow B)\Big] 
\end{eqnarray}
where
$x_b=x_a y_T/(2x_a-y_T)$ and $Q(x) = x\sum_f [q_f(x)+\overline{q}_f(x)]$.
The graphs giving rise to the different hard parts are shown in figure
\ref{graphs}.
Notice that an expression $(xG(x_b,B)+\frac{4}{9}Q(x_b,B))$ appears. This 
so called {\it effective parton distribution} is due to the different values
of the following subprocesses:
\begin{eqnarray}
qq\rightarrow qq ~:~ qg\rightarrow qg ~:~ gg\rightarrow gg
\approx 1 ~:~ C_A/C_F ~:~ (C_A/C_F)^2.
\end{eqnarray}
One therefore is able to describe the whole process in terms of a {\it single effective
subprocess} with an {\it effective} parton distribution
\begin{eqnarray}
Q'(x) = xG(x) + \frac{C_F}{C_A} Q(x) = xG(x)+\frac{C_F}{C_A} x\sum_f [q_f(x)+\overline{q}_f(x)]
\end{eqnarray}
with $C_F/C_A = 4/9$.\\
In the next section we will discuss modifications to the cross section 
due to nuclear effects.
\\
\\
\centerline{\bf 3. Modifications of the cross section: nuclear shadowing}
For the case of prompt photon production in a nucleus nucleus reaction one has to deal with 
{\it nuclear} parton distributions entering the cross section, 
$xG^A(x,Q^2)$ and $F_2^A(x,Q^2)$. It is well known from $eA$, $\mu A$ and $pA$ scattering
\cite{slac,nm,drell} that over the whole
$x$-range, one finds
$f_i^A(x,Q^2) \neq A f_i^P(x,Q^2)$ for the parton densities. Here one has 
nuclear shadowing, anti-shadowing, the EMC effect and Fermi motion, depending on the 
increasing $x$-region \cite{frankfurt}.
In the present case we are mainly 
interested in the small-$x$ region, i.e.~in shadowing. At midrapidity one typically
probes the correlators down to momentum fractions $x\approx k_T \sqrt{s}/(1-k_T \sqrt{s})$. 
Thus in the interesting $k_T$ range at RHIC one reaches values $x\approx 0.01$ and $x\approx 
3.4 \cdot 10^{-4}$ at LHC.\\
The shadowing effect described here is a depletion of the parton densities in the nucleus.
This effect can be understood in terms of parton-parton fusion in the infinite-momentum frame 
\cite{mq}. 
If the longitudinal wavelength 
of a parton exceeds the contracted size of a nucleon (or the inter-nucleon distance) inside 
the Lorentz contracted nucleus, partons
originating from different nucleons can fuse, resulting in a depletion of the 
parton densities at smaller $x$ and in an enhancement at larger $x$. One finds from 
$1/xP \approx 2 R_n M_n /P$ that shadowing shows up at
values $x \approx 0.1$.
A saturation of the shadowing effect can be expected when the longitudinal 
parton wavelength exceeds the size of the nucleus.\\
One also expects a shadowing effect from a transverse overlap of the partons as is
seen in the free proton. Here one finds that, for sufficiently small values of $x$ and/or
$Q^2$, the total transverse area occupied by gluons becomes larger than the transverse
area of the hadron. This happens when $xG(x)\geq Q^2 R^2$, with a transverse parton
size $1/Q^2$ ($Q^2=-q^2$) and a proton radius $R$.
Hence, one finds a depletion of the gluon and sea 
distribution at values $x\leq 0.01$ as shown in figure \ref{MRS-shad}.\\
%
%
It is obvious that shadowing depends on the one hand on the mass number 
of the nucleus and on the other hand on the $x$ and $Q^2$ values. The $Q^2$ dependence of the
ratio $(1/A)f_i^A(x,Q^2)/f_i^P(x,Q^2)$ is determined by modified DGLAP equations 
(first suggested in \cite{glr} and later proven in \cite{mq})
taking the fusion processes into account \cite{kari1}.\\
In the nucleus those fusion processes are due to an enhancement of the fusion of gluon
ladders stemming from independent partons, whereas in the free nucleon, fusion processes from 
non-independent partons dominate.\\
It was shown 
\cite{eskola} that the modifications to the usual DGLAP equations
only affect the ratio by $\approx (6-7) \%$.
Because the detected photons are in a quite narrow range of $k_T$ where
$k_T^2\approx Q^2$, for simplicity, a parametrization for both 
$G^A(x,Q^2)$ and $F_2^A(x,Q^2)$ is used. The first Ansatz \cite{eskola} assumes a similar
size of the shadowing of $F_2^A (x, Q^2)$ and of the gluon distribution at the initial scale
$Q_0=2$ GeV.
The evolution of $F_2(x, Q^2$) is quite moderate, whereas the gluon shadowing vanishes faster with
increasing $Q^2$. This can be understood in terms of the momentum flow into gluons at higher 
$Q^2$ as predicted by the DGLAP equations.\\
The parametrisation of the $(1/A)F_2^A(x,Q^2)/F_2^P(x,Q^2)$ ratio at the
input scale, $Q_0=2$ GeV, given in \cite{eskola}, is shown in figure \ref{fig-kari}.
%
%
Nevertheless, the same parametrisation for both 
distributions is used, so one does not need to solve the modified DGLAP equations for the narrow $k_T$ range. 
It is clear that thereby the shadowing effect is slightly overstimated 
when one tries to be consistent in one Ansatz.\\
Due to the electromagnetical neutrality,
the $R_G(x,Q^2)$-ratio is not accessible with deep inelastic scattering processes.
A second Ansatz \cite{eskola} with a much stronger shadowed
gluon distribution ($R_G(x,Q^2)$ is roughly 45 $\%$ smaller than $R_{F_2}$ at $x=0.001$) 
has been presented at the initial 
value $Q=Q_0 =2$ GeV. The neglect of the $Q^2$ dependence of $R_G(x,Q^2)$
is absorbed by the systematic uncertainty in the Ansatz. 
But there definitely remains an 
overestimate due to
the neglect of the $Q^2$ dependence of $R_{F_2}(x,Q^2)$ which changes by 
$\approx 9\%$ as $Q$ varies from 2 GeV to 5 GeV at $x=0.001$. Therefore,
our overestimate alters the result towards larger $k_T$ by $\approx 10\%$.\\
We show the effect of shadowing for the two energies $\sqrt{s}=200$ AGeV and $\sqrt{s}=5.5$ ATeV.
%
%
The shadowing effect increases as the energy increases which can easily be understood in terms
of the momentum fraction: as one typically probes the correlators (here the parton
distribution functions entering the hand-bag graph) at values $x\approx 2
k_T/ \sqrt{s}$ at $y=0$, this formula also immediatly explains the drop of the ratio towards 
larger transverse momenta. The decrease of the shadowing correction with
increasing $k_T$ can also be understood in terms of the lower integration boundary 
$x^a_{min}$ in Eq. (\ref{cross-section}) which increases with the transverse momentum of the photon.\\
Each cross section is computed on the nucleon-nucleon level as $d\sigma _{pp}/dy$. Then
this quantity is transformed to the number of events per rapidity in a central nucleus nucleus
reaction by multiplying with the nuclear overlap function $T_{AA}(b)$ at
zero impact parameter:
\begin{equation}
\frac{dN_{AA}}{dy}=T_{AA}({\bf b} =0)\frac{d\sigma _{pp}}{dy}~~~~.
\end{equation}
One should not forget that the results for 
{\it central}
collisions are about four times larger than those for collisions {\it
averaged} over all impact 
parameters.\\
Numerically one finds that $T_{AA}({\bf b} =0) \approx A^2/\pi R_A^2$,
\begin{equation}
T_{PbPb} = \frac{32}{{\rm mb}},~~T_{AuAu} = \frac{29}{{\rm mb}}~~~~.
\end{equation}
This procedure is equivalent to neglecting nuclear 
shadowing effects. To explicitly take into account those effects one has to multiply 
each parton distribution with the parametrized ratio $R_{F_2} = (1/A)F_2^A/F_2^N$.\\
\\
\\
\centerline{\bf 4. The hydrodynamical calculations}
For comparison we have also computed the transverse momentum spectrum   
of {\it thermal} hard photons. As a  model for the time-evolution of the
thermalized system we assume a three-dimensional hydrodynamical
expansion with cylindrical symmetry and longitudinal boost invariance
\cite{Bj}. For the RHIC energy we employ an initial temperature of
$T_i=533$~MeV and an initial time of $\tau_i=0.1236$~fm/c, while
for LHC we assume $T_i=880$~MeV and $\tau_i=0.1$~fm/c.
The resulting final pion
multiplicity in central Au+Au reactions (i.e.\ initial transverse QGP radius
$R_T=6.5$~fm) is $dN^\pi/dy\approx1460$ at RHIC and
$dN^\pi/dy\approx5300$ at LHC, if the hydrodynamical
evolution is isentropic and if the QGP and the hadron gas consist of
$u$, $d$ quarks, gluons and pions, respectively \cite{hwa}. According to   
present knowledge
\cite{T0}, from the viewpoint of maximum thermal radiation these initial
conditions (and the assumption of full thermal and chemical equilibrium)
have to be considered as the most favorable ones. Therefore, 
we have also computed the thermal photon spectrum for a less rapid   
thermalization and a lower initial temperature ($\tau_i=0.5$~fm/c,
$T_i=300$~MeV, $dN^\pi/dy\approx1050$ for RHIC and $\tau_i=0.25$~fm/c,
$T_i=650$~MeV, $dN^\pi/dy\approx5350$ for LHC).
We assume a freeze-out temperature of $T_f=100$~MeV.
The photon spectrum above $k_T=2$~GeV, however, does not depend sensitively on
this number, except in the case of slow thermalization and low initial 
temperature ($\tau _i = 0.5$~fm/c, $T_i=300$~MeV). 

The equation of state is that of
an ideal gas of massive $\pi$, $\eta$, $\rho$, and $\omega$ mesons below
$T_C=160$~MeV. For $T>T_C$ we assume an ideal QGP (massless, noninteracting
$u$, $d$ quarks and gluons) described within the MIT bag-model.
The bag constant is chosen such that the pressures of the two phases match
at $T=T_C$ ($B^{1/4}=235$~MeV) thus leading to a first order phase transition.

\begin{figure}
\centerline{\psfig{figure=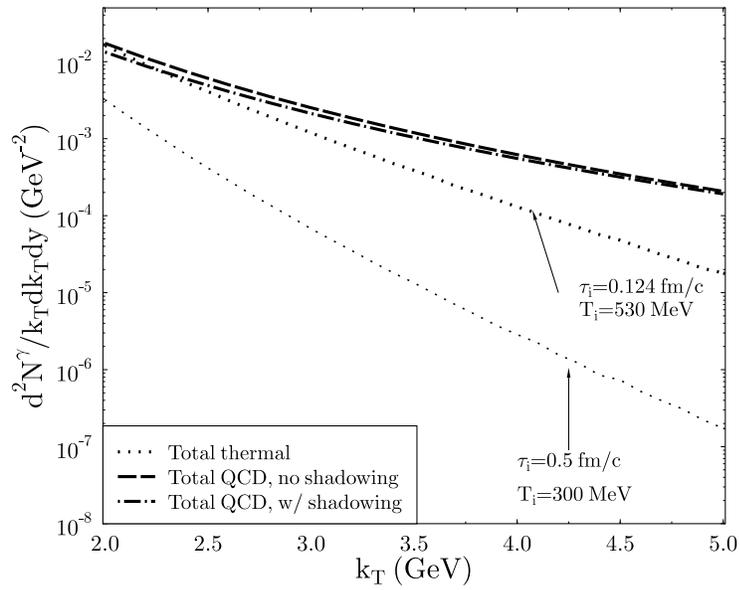,width=18cm}}
\caption[]{\sf Transverse momentum distribution of prompt and thermal photons produced in a 
central Au+Au collision at $\sqrt{s}=200$ AGeV.}
\label{pptogamma1}
\end{figure}

\begin{figure}
\centerline{\psfig{figure=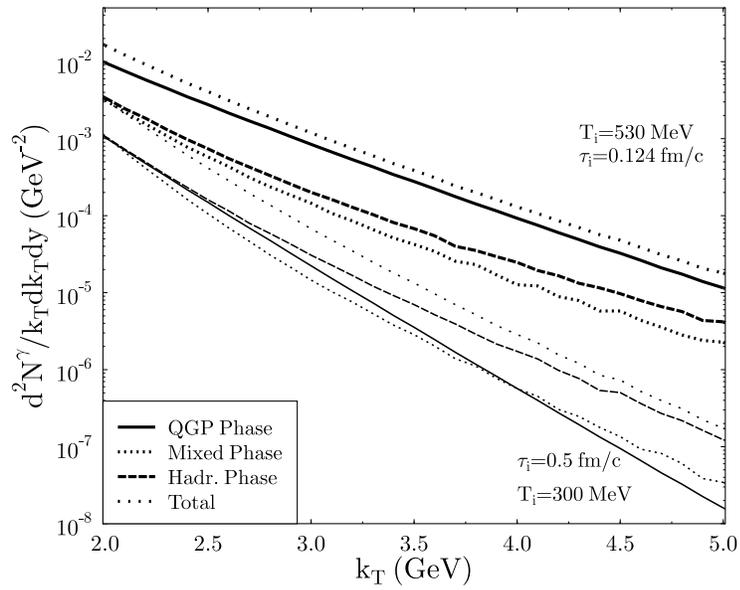,width=18cm}}
\caption[]{\sf Transverse momentum distribution of thermal photons produced in the different phases
at $\sqrt{s}=200$ AGeV with different initial conditions.}
\label{adrian1}
\end{figure}

\begin{figure}
\centerline{\psfig{figure=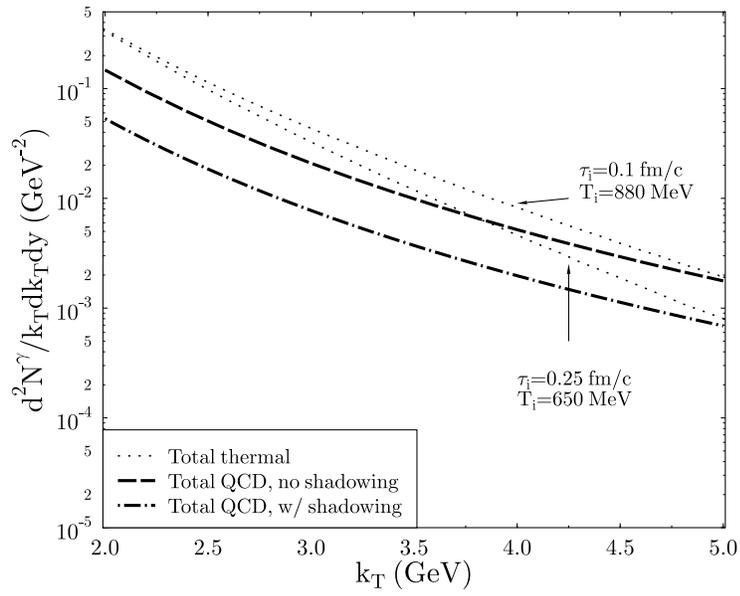,width=18cm}}
\caption[]{\sf Same as \ref{pptogamma1} but for $\sqrt{s}=5.5$ ATeV in Pb+Pb.}
\label{pptogamma1a}
\end{figure}

\begin{figure}
\centerline{\psfig{figure=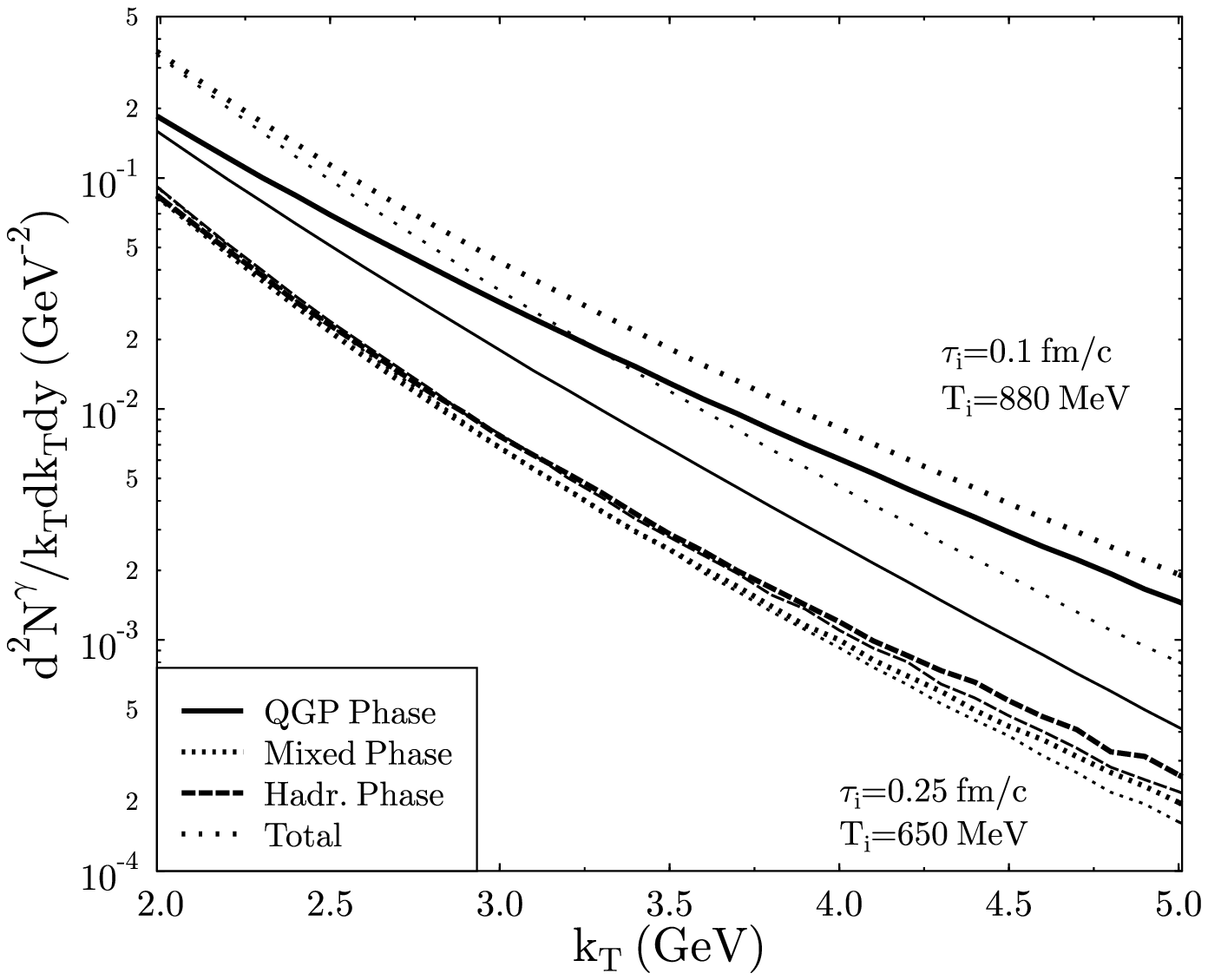,width=18cm}}
\caption[]{\sf Same as \ref{adrian1} but for $\sqrt{s}=5.5$ ATeV.}
\label{adrian2}
\end{figure}

The number of emitted direct photons per infinitesimal space-time volume
in each of the three phases is parametrized as \cite{KLS}
\begin{equation}
\label{rate}
E\frac{dN^\gamma}{d^4x\,d^3k} = \frac{5\alpha \alpha_S}{18\pi^2}  
T^2 e^{-E/T} \ln \left( \frac{2.912E}{g^2 T} +1 \right)   \quad.
\end{equation}
$E$ is the photon energy in the local rest frame. In our hydro calculations
we fix $\alpha_s=g^2/4\pi=0.3$. A logarithmic temperature
dependence of $\alpha_s$ does not alter the results significantly.
Eq.\ (\ref{rate}) accounts for
pion annihilation ($\pi\pi\rightarrow \rho\gamma$), and Compton-like
scattering ($\pi\rho\rightarrow \pi\gamma$, $\pi\eta\rightarrow \pi\gamma$)
off a $\rho$ or $\eta$ meson (in lowest order perturbation theory). In the
QGP (2 massless flavors), alternatively, quark-antiquark annihilation
($q\overline{q}
\rightarrow g\gamma$) and Compton-like scattering off a gluon
($q,\overline{q}+g\rightarrow q,\overline{q}+\gamma$) are considered.
In view of the uncertainties of the initial conditions and the time evolution
of the temperature, which enters the thermal photon production rate
exponentially, we have not attempted to include higher order corrections,
hadronic formfactors, or additional processes (e.g.\ $a_1\rightarrow  
\pi\gamma$ decays etc.). Finally, the thermal photons produced at each
individual space-time point are summed incoherently to obtain the spectrum
of photons emitted in a nuclear collision.\\
\\
\\
\centerline{\bf 5. Results}
We have calculated the transverse momentum spectrum (at midrapidity)
of prompt photons, $d^2 N^{\gamma}/k_T dk_T dy$ produced in the very early
phase ($1/k_0 \approx 0.1$ fm/c) of an ultra-relativistic heavy ion collision via
perturbative QCD. We take
the transverse photon momenta in the range $k_T=2...5$ GeV. The lower bound is chosen such that
perturbation theory is still valid
and one does not have to take into account matrix elements of higher twist,
such as $\left< P | F^{+\alpha}\overline{\psi} \gamma ^{+} F^{+}_{\alpha}\psi | P \right>$.
We use a rather moderate estimate for
the higher order terms, namely K=1.5 for RHIC and K=1.0 for LHC.
We compare those prompt photons to the thermal radiation, calculated as outlined above
(cf. figures \ref{pptogamma1} and \ref{pptogamma1a}).
In the relevant range we find that the photons from Bremsstrahlung dominate over
those coming from the QCD Compton process by a factor $\approx 1.6[ {\rm ln} (1/x_T)-1.1]$ which 
itself dominates over the annihilation
process as also found in \cite{alam} and \cite{hwa}. In our calculation we 
find that the Bremsstrahlung contribution at RHIC lies above the Compton contribution
in the whole $k_T$-range whereas in \cite{alam} one had a cross-over point 
at $k_T \approx 3.5$ GeV. This different behavior should be due to the parton distributions
(Duke/Owens set I) used in \cite{alam}.
Also, for the parton distributions employed by us, prompt photons dominate
over the thermal radiation in the entire transverse momentum range at RHIC, cf. 
figure \ref{pptogamma1}. 
%
%
%
The inverse slope $1/T = - d/dk_T ~{\rm ln} ~(dN/k_T dk_T dy)$ of the prompt QCD-photons in 
the interval $k_T=2...5$ GeV is
$\approx$ 500-1000 MeV at RHIC and 500-1100 MeV at LHC, depending on $k_T$.
Hydrodynamics predicts that the inverse slope of the thermal photons can hardly exceed
500 MeV at RHIC and 700 MeV at LHC, even if very rapid thermalization 
times $\tau _i \approx 1/k_0$ and high initial temperatures $T_i \approx 3\left< E_T\right>$
are assumed.
The effect of the nuclear shadowing correction is quite moderate at the RHIC energy
as also found in minijet calculations \cite{eskola}.
Note also (figure \ref{adrian1}) that for the second set of initial conditions the hadronic phase
dominates the thermal photon spectrum, due to the strong radial flow. A higher freeze-out 
temperature would, however, reduce this contribution.\\
The effect of nuclear shadowing increases for higher
energies (LHC, see figure \ref{pptogamma1a}) and yields corrections up to a factor of $\approx$ 3.5, and therefore
is not negligible, especially at smaller momentum fractions.
Without shadowing corrections one has a cross over point with the thermal yield ($\tau _i =
0.25$ fm, $T_i = 650$ MeV) at $k_T \approx$ 3.75 GeV. When taking into account
the shadowing corrections one finds the cross over point at larger transverse
momenta ($k_T \approx 5$ GeV). For the higher initial temperature of $T_i = 880$ MeV,
the cross over point is shifted to even larger values of $k_T$.\\
Thus, even for the lower  initial temperature and larger
formation time, photons from primary QCD processes in the important
range $k_T = 2...5$ GeV are negligible at LHC if shadowing is taken into account.
Also, for the initial conditions expected at LHC, the thermal spectrum is dominated by 
photons produced in the QGP phase (figure \ref{adrian2}). 
%
%
%
\\
\\
\\
\centerline{\bf 6. Summary and conclusions}
In this paper we have computed the number of prompt photons with 2 GeV $\leq k_T\leq 5$ GeV 
in a central $AA$ collision at energies of $\sqrt{s}=200$ AGeV and $\sqrt{s}=5.5$ ATeV at midrapidity 
to study nuclear modifications of the naive
extrapolations from pp results.
We investigated different sources for the photons
and also discussed thermal photon production within a hydrodynamical 
model and various sets of initial conditions.
We used the nuclear overlap function to derive the multiplicities from the cross sections
and assumed central collisions ($\bf b$=0) which yield multiplicities about 
four times larger than those resulting from impact parameter averaged 
calculations.\\
With an upper bound for nuclear shadowing and a relatively conservative estimate for the 
higher orders, expressed in the K-factor, at RHIC we obtain multiplicities larger than those 
from the thermalized stage of the reaction. For the LHC energy of $\sqrt{s}=5.5$ ATeV we find that 
the cross over point between the thermal photons and the primary photons significantly
changes towards larger transverse momenta when shadowing corrections are taken into 
account. Even if we slightly overestimate the shadowing effect, one can conclude that 
at LHC the background from QCD processes of the initial nucleons is of minor importance
in the range $k_T = 2...5$ GeV but will still dominate for the very high transverse 
photon momenta.\\
The inverse slope of the QCD-photons ($\approx$ 500-1000 MeV) significantly exceeds
that of the thermal photons (300-500 MeV) at RHIC and is slightly larger for
LHC energies. 
\newpage
\def\pr{{\sl Phys. Rev.}~}
\def\prl{{\sl Phys. Rev. Lett.}~}
\def\pl{{\sl Phys. Lett.}~}
\def\np{{\sl Nucl. Phys.}~}
\def\zp{{\sl Z. Phys.}~}

\newpage
\vspace{2cm}
\vspace{2cm}
\begin{figure}
\centerline{\psfig{figure=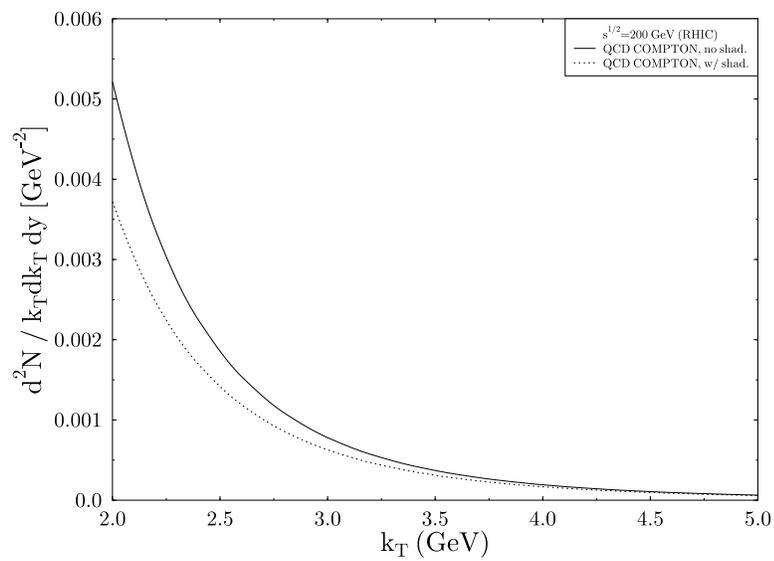,width=14cm}}
\caption[]{\sf QCD Compton photon multiplicity distribution with and without
nuclear shadowing for $\sqrt{s}=200$ AGeV.}
\label{shadow-rhic}
\end{figure}
\begin{figure}
\centerline{\psfig{figure=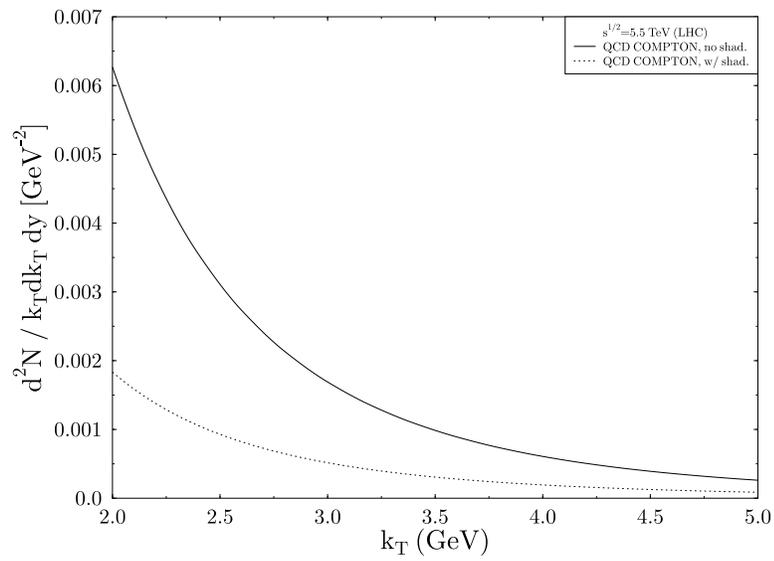,width=14cm}}
\caption[]{\sf Same as figure \ref{shadow-rhic} but for $\sqrt{s}=5.5$ ATeV.}
\label{shadow-lhc}
\end{figure}

\end{document}